\documentstyle[prb,aps,preprint,psfig]{revtex}
\begin{document}
\draft
\title{Negative Differential Resistance in the Scanning Tunneling 
Spectroscopy of Organic Molecules}
\author{Yongqiang Xue and Supriyo Datta}
\address{School of Electrical and Computer Engineering, Purdue University, 
West Lafayette, IN 47907}
\author{Seunghun Hong and R. Reifenberger}
\address{Department of Physics, Purdue University, West Lafayette,
IN 47907}
\author{Jason I. Henderson and Clifford P. Kubiak}
\address{Department of Chemistry and Biochemistry, University of 
California, San Diego, La Jolla, CA 92093} 
\date{\today}
\maketitle
\begin{abstract}
The conductance-voltage spectrum of molecular nanostructures measured by 
scanning tunneling spectroscopy (STS) is generally assumed to 
reflect the local density of states of the molecule. This excludes the 
possibility of observing negative differential resistance (NDR). We report 
here the observation of NDR in the scanning tunneling microscope (STM) 
current-voltage (I-V) characteristics of self-assembled monolayer (SAM) of 
4-p-Terphenylthiol molecules on gold substrate measured using a platinum 
probe. We argue that the NDR arises from narrow structures in the local 
density of states at the tip apex atom and show that depending on the 
electrostatic potential profile across the system, NDR could be observed 
in one or both bias directions. 
\end{abstract}
\begin{figure}
\centerline{\psfig{figure=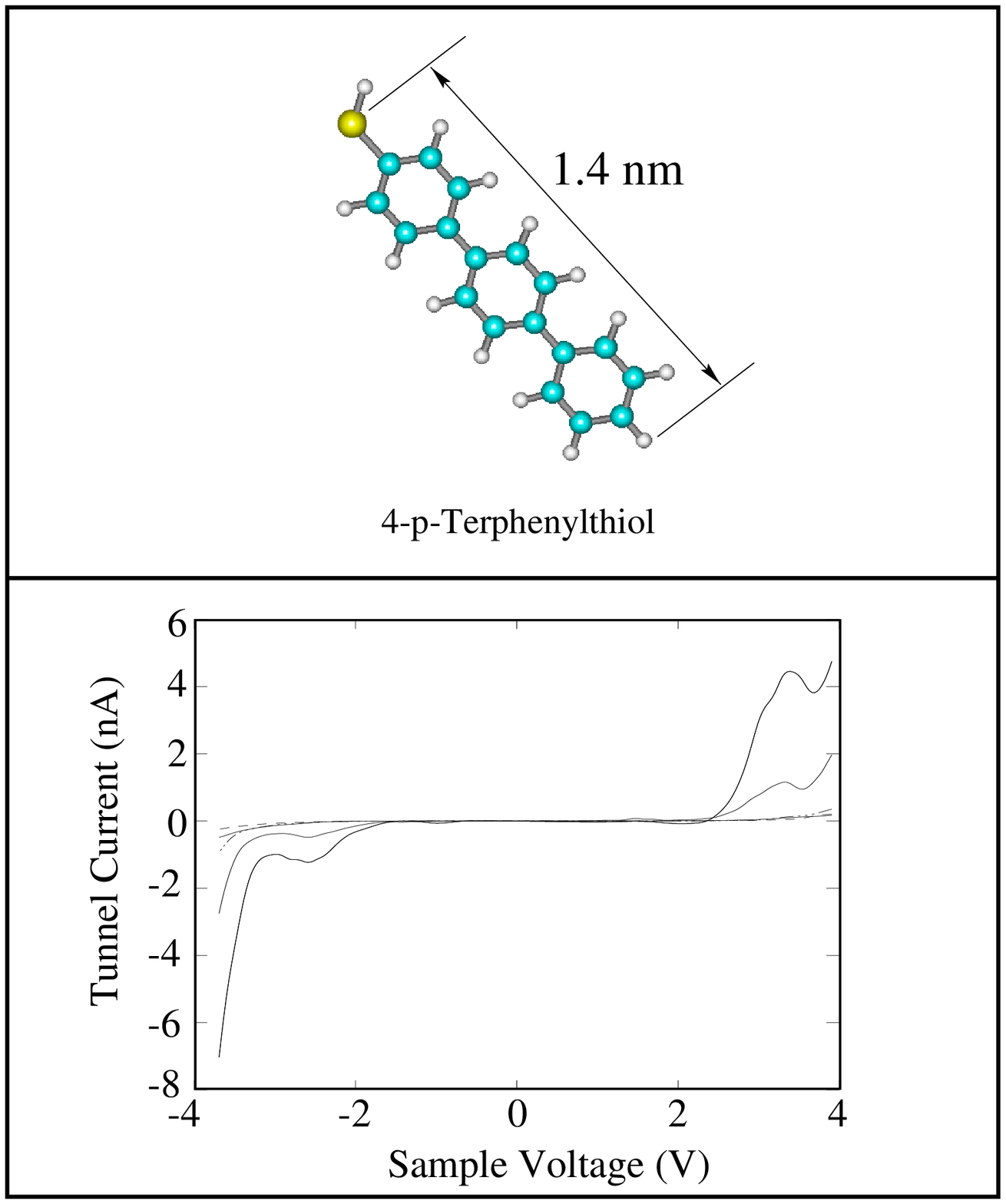,angle=0.,height=5.0in,width=5.0in}} 
\vspace{0.1cm}
\caption{Experimental I-V characteristics of the molecule as
a function of the molecule-tip distance. A negative differential 
resistance is observed for small tip/SAM separation. The set point voltage 
is -5V; the set point currents range between 3.05 nA and 30.5 nA.}
\label{xueFig1}
\end{figure}

Electron transport through molecular nanostructures has been widely 
studied in recent years, using either the scanning tunneling microscope 
or mechanically controllable break junctions\cite{datta,reed}. The 
conductance-voltage spectrum, $dI/dV$, is commonly assumed to reflect 
the local density of states (LDOS) of the molecule\cite{tersoff}: 
$dI/dV \sim \rho(E=E_{F}+eV)$, $E_{F}$ being the equilibrium 
Fermi energy. A more accurate description of the conductance-voltage 
spectrum, as shown in Ref.\onlinecite{datta}, is to take a weighted 
average of the density of states: 
\begin{equation} 
 dI/dV \sim \eta \rho(E=\mu_{1})+(1-\eta)\rho(E=\mu_{2})
\label{G-V}
\end{equation} 
Here $\mu_{1}$ and $\mu_{2}$ are the electrochemical potentials in the 
two contacts and the factor $\eta$ describes the voltage division across 
the molecule: $\mu_{1}=E_{F}-\eta eV$ and $\mu_{2}=E_{F}+(1-\eta)eV$.  
Based on this viewpoint, NDR, i.e., a negative slope in the I-V curve, 
cannot occur since the density of states is nonnegative. Experimentally, 
however, we often observe negative differential resistance for monolayer 
of long molecules self-assembled on gold substrate such as the one shown 
in Fig.\ \ref{xueFig1}.

\begin{figure}
\centerline{\psfig{figure=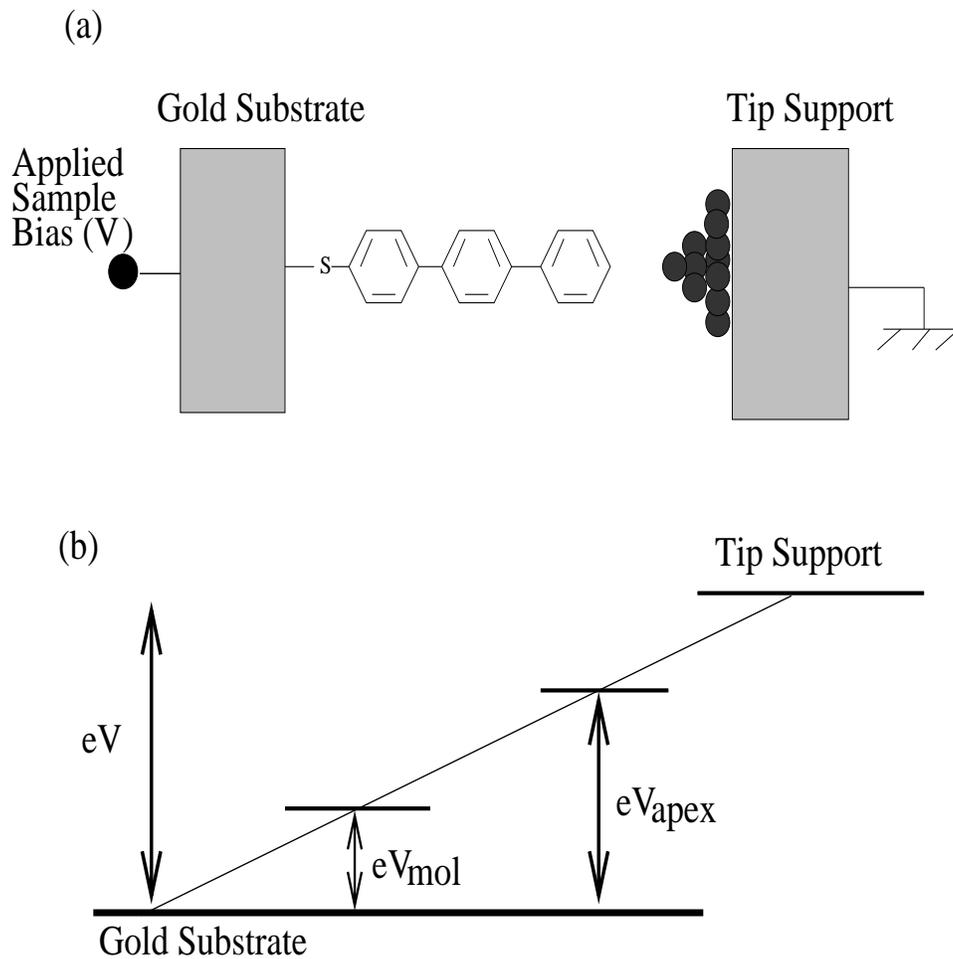,angle=0.,height=5.0in,width=5.0in}}
\vspace{0.1cm}
\caption{{\bf (a)} SAM of 4-p-Terphenylthiol on gold substrate (only one 
molecule is shown here for clarity), also shown is the STM tip. {\bf (b)} 
Electrostatic potential profile under applied bias, here we only show the 
case of positive sample bias.} 
\label{xueFig2}
\end{figure}

A possible scenario that can lead to NDR at the atomic level was studied 
theoretically by Lang\cite{lang} and experimentally by Avouris and 
co-workers\cite{Avouris}. The essential argument is that if there is a 
weak link between two parts of the conducting system, each of which has 
relatively narrow features in the density of states in the energy range of 
interest, NDR is likely to occur. This is readily understood from the 
transfer Hamiltonian point of view\cite{W&G}, which relates current to 
the product of the density of states $\rho_{L}$ and $\rho_{R}$ on both 
sides of the weak link. As the bias is changed, current can decrease if 
two narrow structures in $\rho_{L}$ and $\rho_{R}$ move away from 
alignment. For the structure we studied (Fig.\ \ref{xueFig2}), the weak 
link is the STM tip-molecule junction. If the tip has a featureless 
density of states, as is implicitly assumed in the derivation of Eq.\ 
(\ref{G-V}), we wouldn't expect this scenario to apply. However, narrow 
features in the density of states can develop at the tip atoms in a 
realistic model of tip. This has been widely recognized in the STM study 
of surfaces\cite{W&G} and is used to explain the NDR in STM I-V 
characteristics of boron-exposed silicon surface\cite{Avouris}. More 
recently, Yeyati et al.\ \cite{yeyati} have studied the electronic 
structure of a sharp gold tip in the context of conductance quantization 
in gold atomic-size contacts and shown that narrow resonant states can 
develop at the tip apex atom provided the tip geometry is sufficiently 
sharp.

\begin{figure}
\centerline{\psfig{figure=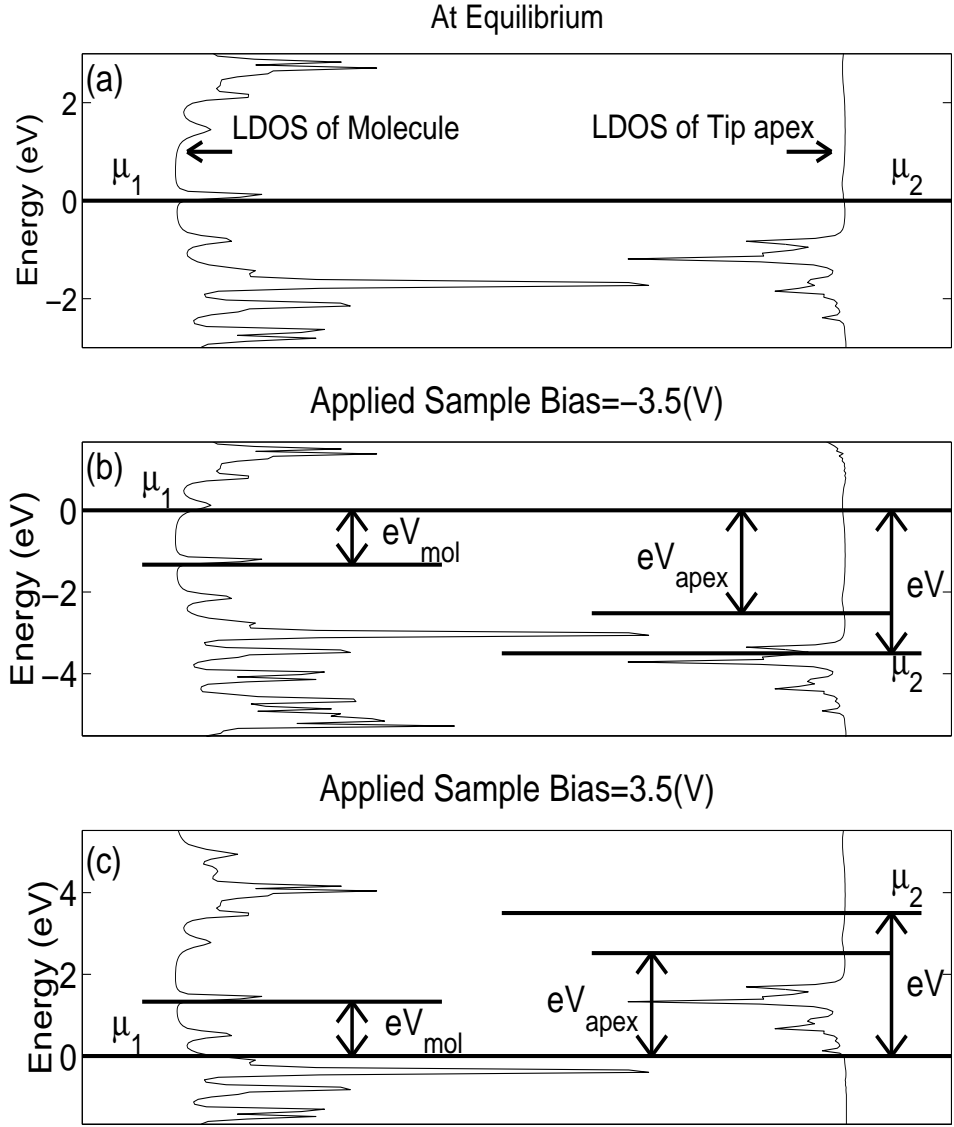,angle=0.,height=6.0in,width=5.0in}}
\vspace{0.1cm}
\caption{Illustration of how the LDOS structures of molecule and tip apex 
atom sweep past each other under applied bias. The electrochemical 
potential of gold substrate $\mu_{1}$ is taken as energy reference (The 
LDOS curves have been horizontally offset for clarity). Parameters used 
are the same as those used for the I-V calculation in Fig.\ \ref{xueFig4}
(a). {\bf (a)} At equilibrium, $\mu_{1}=\mu_{2}=E_{F}$; {\bf (b)} At 
negative sample bias, the LDOS curves of molecule and tip apex atom float 
down relative to gold substrate; {\bf (c)} At positive sample bias, the 
LDOS curves of molecule and tip apex atom float up relative to gold 
substrate. NDR already occurs before sample bias reaches $\pm3.5$(V).} 
\label{xueFig3}
\end{figure}

In this paper we explain the occurrence of NDR by taking into account the 
electronic structure of sharp platinum tip (used in our STM measurement
\cite{note1}) as well as the electrostatic potential profile across the 
tip/molecule system. Using a tip model similar to Yeyati et al.\ 
\cite{yeyati}, we find narrow structures in the LDOS of tip apex atom 
below the equilibrium Fermi energy $E_{F} $ (Fig.\ \ref{xueFig3}(a)). 
Since the LDOS of molecule also exhibits sharp structures, we expect NDR 
to occur under applied bias. However, since the narrow structures of LDOS 
at tip apex atom are below $E_{F}$, NDR will occur only at positive sample 
bias if the electrostatic potential at the tip apex atom is the same 
as that of the tip support. But NDR can occur in both bias directions if a 
significant amount of the voltage is dropped between the tip apex atom and 
the tip support. We argue that for very sharp tip geometry, this is likely 
to be true when the tip is close to the molecule. Since the tip is 
composed of a small cluster of platinum atoms, the screening length in 
this region can be much larger than that in the bulk and also larger than 
the cluster size, making it possible to maintain an electrostatic 
potential drop between the tip apex and the tip support under applied 
bias. Experimentally, we observe NDR in both bias directions when the tip 
is very close to the molecule, in agreement with the theory.

Our model is illustrated in Fig.\ \ref{xueFig2}(a). The self-assembled 
monolayer (SAM) of 4-p-Terphenylthiol molecules are synthesized using 
standard procedure. I-V data presented here (Fig.\ \ref{xueFig1}) 
represents the average of 25-50 consecutive I(V) sweeps taken at a fixed 
position on the sample\cite{hong}. The SAMs attach strongly to the gold 
$(111)$ surface through the sulfur end atom forming a strong chemical bond 
with good orbital overlap\cite{xia}. The STM tip usually couples weakly to 
the molecule, corresponding to the physisorption situation. As in our 
previous studies\cite{datta}, we'll use the Extended H\"{u}ckel Theory 
(EHT) to describe the whole molecule-STM system (for a recent 
justification of using EHT in STM study of metal surfaces, see Ref.\ 
\onlinecite{cerda}), taking into account the $5d6s6p$ orbitals of 
platinum and gold.

Similar to Yeyati et al.\ \cite{yeyati}, we model the tip geometry as a 
small cluster of Pt atoms stacked on the $\langle 111 \rangle$ surface of 
semi-infinite support\cite{binh}, originating from a monatomic apex, 
adding 3 nearest neighbor atoms to the 2nd layer and 7 nearest neighbor 
atoms to the 3rd layer (we also include 14 nearest neighbor atoms on the 
surface of the tip support and 6 atoms on the gold substrate surface in 
the calculation of I-V characteristics). The on-site energies for the 
orbitals of the 3 layers of tip atoms are modified self-consistently by 
adjusting the occupation number of each orbital to that of the neutral 
atom until local charge neutrality is achieved on each atomic site\cite
{yeyati,cerda,Book3}.
 
We calculate current using\cite{Datta}:
\begin{eqnarray}
 I &=& \frac{2e}{h} \int_{-\infty}^{+\infty}dET(E,V)
[f(E-\mu_{2})-f(E-\mu_{1})] \nonumber \\
  &\simeq& \frac{2e}{h} \int_{\mu_{1}}^{\mu_{2}}dET(E,V)
\label{scattering}
\end{eqnarray}
where $f(E)$ is the Fermi distribution and $\mu_{1}$, $\mu_{2}$ are the 
electrochemical potential of the gold substrate and tip support 
respectively, $\mu_{2}=\mu_{1}+eV$\cite{note2}. The transmission $T(E,V)$ 
can be calculated using the scattering theory of transport, as we shall 
describe shortly, but we can get more insight if we use the transfer 
Hamiltonian formalism\cite{W&G} to relate the transmission  to the local 
density of states on either side of the STM tip-molecule junction:
\begin{equation}
 T(E,V)=4 \pi^{2} \mid M_{LR} \mid ^{2} \rho_{L}(E-eV_{L}) 
        \rho_{R}(E-eV_{R})
\label{T:Tunneling}
\end{equation}
where $M_{LR}$ is the coupling matrix element and $V_{L}$ and $V_{R}$ are 
the electrostatic potentials of the molecule and the tip apex atom 
respectively.

The LDOS of both the molecule and the tip apex atom show narrow features 
(Fig.\ \ref{xueFig3}(a)), which are calculated from the Green's function 
of molecule-tip system using\cite{sancho}: 
\begin{equation}
  \rho_{molecule(tip)}=-\frac{1}{\pi} Tr\{Im(GS)_{molecule(tip)}\}
\end{equation}
where $G(E)$ is defined by $G(E)(ES-H)=(ES-H)G(E)=I$, and $S$ is the 
overlap matrix. The narrow structures of LDOS at tip apex can be 
understood qualitatively from the fact that platinum is third row 
transition metal, whose LDOS around the equilibrium Fermi energy $E_{F}$ 
is mostly due to the contribution of the 5d orbitals, located slightly 
below $E_{F}$. For the given sharp tip geometry, the coupling of the tip 
apex atom to its local environment is much weaker comparing to that in 
the bulk, so the levels remain fairly sharp. 

\emph{What is the electrostatic potential profile?} Applying an external 
bias changes the relative electrochemical potential of the gold substrate 
and tip support, which are taken as infinite electron reservoirs. The 
electrostatic potential $\varphi(r)$ is determined self-consistently from 
the Poisson equation $\nabla^{2} \delta \varphi(r)=e\delta n(r)$. We only 
need to calculate the change in electrostatic potential 
$\delta \varphi(r)$ (which in turn modifies the molecular Hamiltonian) 
since its equilibrium value has been included in the equilibrium 
Hamiltonian. The electrostatic potential change in each electrode follows 
that of the electrochemical potential which provides boundary condition 
to the Poisson equation. The important point is that we take the 3 layers 
of tip atoms and the molecule on an equal footing, viewing the whole 
molecule-tip system as an ``extended molecule'' sandwiched between the two 
electrodes. In the region near the tip apex, both the electrochemical and 
the electrostatic potential can be different from those inside the tip 
support and from each other\cite{mclennan}.

As a first approximation, if we neglect any charge buildup within the 
molecule-tip system, and assume that the gold substrate and tip support 
act as two infinite parallel plates of a capacitor, then the electrostatic 
potential varies linearly, as shown in Fig.\ \ref{xueFig2}(b). For 
simplicity, we assume that the molecular energy levels simply float up by 
an amount around the average electrostatic potential change, that is, we 
neglect any Stark shift of the energy levels due to the electric field 
inside the molecule\cite{note4}. The origin of NDR can be best understood 
by examining how the LDOS structures of the molecule and the tip apex atom 
sweep past each other under applied bias (Fig.\ \ref{xueFig3}). NDR will 
occur for both bias directions if we assume a significant voltage drop 
between the tip apex and the tip support, i.e., $eV_{apex} < eV$, as 
evident from Fig.\ \ref{xueFig3}.

\begin{figure}
\centerline{\psfig{figure=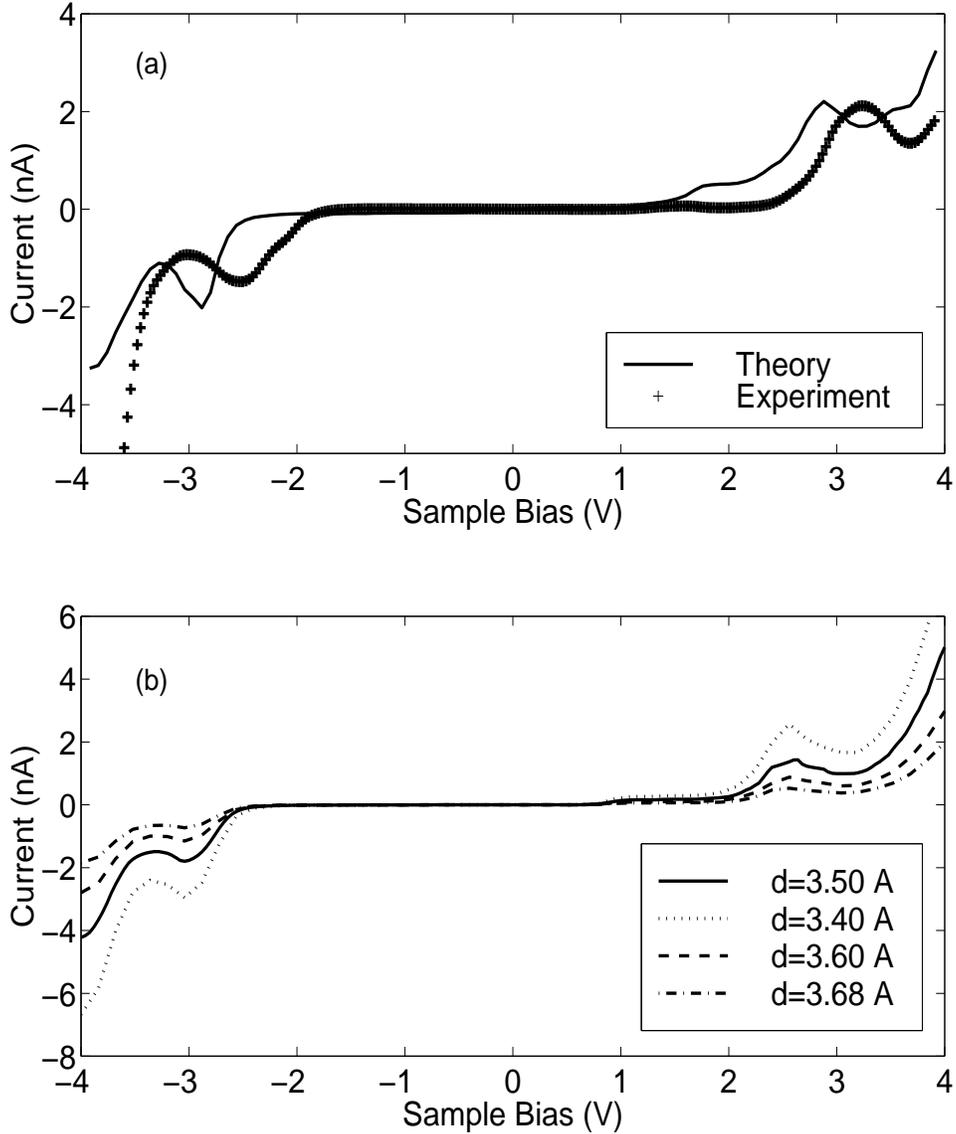,angle=0.,height=6.0in,width=5.0in}}
\vspace{0.1cm}
\caption{ {\bf (a)} I-V characteristics calculated using transfer 
Hamiltonian theory for tip-molecule distance$=3.50$\AA. Parameters used 
are: $\mid M_{LR} \mid ^{2}=8.2\times10^{-8}(eV)^{2}$, $E_{F}=-11.05$(eV) 
and the electrostatic potential change of the molecule is taken as 1.15 
times the average electrostatic potential change. {\bf (b)} I-V 
characteristics calculated using Eq.\ (\ref{Scattering}) for various 
tip-molecule distance d (in unit of angstrom), here we use $E_{F}=-11.15$
(eV), $U=0.1$(eV).}
\label{xueFig4}
\end{figure}

It can be seen from Fig.\ \ref{xueFig4}(a) that this approach predicts the 
I-V characteristics quite well with four fitting parameters, namely 
the tip-molecule distance, the equilibrium Fermi energy $E_{F}$, the 
constant coupling matrix element $M_{LR}$ and the electrostatic potential 
change of the molecule. We also calculated the transmission function using 
the equation (see page $148$ in Ref.\ \onlinecite{Datta}):
\begin{equation}
 T(E,V)=Tr\{\Gamma_{L}(E)G^{R}(E,V)\Gamma_{R}(E-eV)G^{A}(E,V)\}
\label{Scattering}
\end{equation}
which goes beyond the transfer Hamiltonian formalism and takes arbitrary 
sample/electrode coupling into account. Also, a better approximation to 
the electrostatic potential should take charging effects within the 
molecule into account. In tight-binding theory, this gives 
$\delta \varphi_{mol}=V_{0}+U\delta n_{mol}/e$, where $V_{0}$ is the 
average elctrostatic potential change of the molecule without charge 
buildup and $\delta n_{mol}$ is the excess number of electrons in the 
molecule induced by the applied bias. Since the molecule is strongly 
coupled to the gold substrate, we can assume that it is in equilibrium 
with the gold substrate and get $\delta n_{mol}$ by solving the following 
self-consistent equation:
\begin{eqnarray}
\delta n_{mol}  &=&  \int_{-\infty}^{\mu_{1}}\rho_{mol}(E-e\delta 
\varphi_{mol})dE-\int_{-\infty}^{\mu_{1}}\rho_{mol}(E)dE \nonumber \\
&=& \int_{\mu_{1}}^{\mu_{1}-eV_{0}-U\delta n_{mol}}\rho_{mol}(E)dE
\label{Electrostatic}
\end{eqnarray}
The charging energy U can be estimated from the energy gap in molecular 
optical spectra, electron affinity and ionization potential of the 
molecule: $U=(I-A-E_{g})/2-W_{im}$, where $E_{g}$ is the optical gap and 
$W_{im} \simeq \frac{e^{2}}{4R_{mol}}$ is the image potential due to the 
substrate \cite{Book3}($R_{mol}$ is the distance from the 
center-of-mass of molecule to the substrate). Since $I-A \sim4$eV, 
$E_{g} \sim 2.5$eV, and $R_{mol}\approx 8.5$\AA, $U$ is $\sim 0.3$eV. 
We have used $U=0.1$eV in our calculation. The resulting electrostatic 
potential change is included in the calculation of the Green's function 
and transmission probabilities using Eq.\ (\ref{Scattering}). Results of 
such calculation show similar qualitative features to that obtained from 
the simple transfer Hamiltonian argument (Fig.\ \ref{xueFig4}(b)). 

In conclusion, we have presented a simple explanation of the NDR in STM 
I-V characteristics of self-assembled monolayer of 4-p-Terphenylthiol 
molecules in terms of the electronic structure of sharp platinum tip. The 
major approximation in our theory is the use of the Extended H\"{u}ckel 
theory and a simplified treatment of electrostatic potential variation. A 
central factor in understanding the NDR is the narrow structures in the 
LDOS at the tip apex which has been noted by other authors in different 
contexts. However, this alone is not enough to explain the occurrence of 
NDR in both bias directions. NDR for negative sample bias can be 
understood only if we allow for an electrostatic potential drop between 
the tip apex and the tip support. This is possible for very sharp tip 
geometries, where screening length is large. But if the tip is relatively 
flat, the potential drop between the tip apex atom and the tip support 
will be reduced. Indeed, experimentally we do observe NDR only in positive 
sample bias direction for some of our tips. However, even with these tips, 
we observe NDR in both bias directions when we move the tips closer to the 
molecule, indicating a significant increase in the potential drop between 
the tip apex and the tip support. It's not clear to us why the potential 
drop should increase significantly, since the distance range by which the 
tips move is relatively narrow.  Perhaps a more complete theory is needed 
that involves self-consistent evaluation of the electronic structure for 
the entire electrode/sample system, such as described by Lang\cite{lang2}. 
Our main purpose here, is then to show that the electrostatic potential 
profiles play a crucial role in determining the shape of I-V 
characteristics of molecular nanostructures--- a role that has not been 
adequately recognized.

This work was partially funded by the National Science Foundation under
Grant No.\ 9708107-DMR.  One of us (R.R.) would like to thank Norton Lang 
for a stimulating discussion during the very early stages of this work.

\begin{references}
\bibitem{note1} Technically, Pt/Ir tips with composition 80:20 were used 
(Ir atoms have little effect on the tip electronic structure).
\bibitem{note2} Under zero bias, both $\mu_{1}$ and $\mu_{2}$ are equal to 
the equilibrium Fermi energy of the STM-molecule system $E_{F}$ which can 
be determined in principle using the relation $N=2 \times \sum_{i} 
\frac{1}{\pi} \arctan{\frac{\Gamma_{i}}{\epsilon_{i}-E_{F}}}$ (N is the 
number of electrons in the molecule, see Ref.\ 1). However, 
the resulting $E_{F}-N$ curve shows quick variation in the region around 
the neutral case, making it better to use $E_{F}$ as a fitting parameter 
in this region (experimentally the charge transferred between the contacts 
and the molecule at equilibrium is usually much less than one electron).
\bibitem{note4} In general, the molecule will be polarized by an external 
electric field, whose polarization can be calculated using a 
self-consistent procedure such as density functional theory. This effect 
will reduce the electric field within the molecule and correspondingly, 
giving rise to a larger potential drop across the tip than that obtained 
from the assumption of linear potential variation.
\end{references}
\end{document}